\documentclass[letterpaper]{article}
\usepackage{aaai2026}
\usepackage{times}
\usepackage{helvet}
\usepackage{courier}
\usepackage[hyphens]{url}
\usepackage{graphicx}
\usepackage{booktabs}
\usepackage{amsmath}
\usepackage{amssymb}
\usepackage{natbib}
\nocopyright
\title{Informational Help-Seeking on Reddit Did Not Decline After ChatGPT}
\author{Hazem Ibrahim\textsuperscript{\rm 1,*}, Yasir Zaki\textsuperscript{\rm 1}}
\affiliations{\textsuperscript{\rm 1}Computer Science, Science Division, New York University Abu Dhabi, UAE\\
\textsuperscript{*}Corresponding author: hazem.ibrahim@nyu.edu}

\begin{document}
\maketitle

\begin{abstract}
Did people stop asking other people for advice online once generative AI could answer their questions? Prior work on ChatGPT's effect on online help-seeking disagrees in both size and sign, in part because no study has compared affected communities against similar communities that AI cannot easily substitute for, over the same months. In this paper, we track monthly post counts in 26 Reddit informational communities against 90 size-comparable hobby communities over the same six calendar months before and after the launch of ChatGPT. We also repeat the entire analysis at 66 earlier dates, before ChatGPT existed, to see what our method reports when no ChatGPT-effect exists. We find that informational help-seeking did not decline. Our results rule out any decline in posting larger than 3.4\%, far smaller than the 8\% to 25\% declines documented in prior work. Steady post counts could still be misleading if AI-written posts had replaced human ones. We test this possibility by scoring 274{,}411 posts and 223{,}775 comments with AI-text detectors, compared in a way that cancels out detector false-positives on human-written text. AI-written posts rose only 2--3 percentage points more in informational communities than in hobby communities, short of the 5.1 points that would be needed to hide even the smallest decline previously reported for Reddit. In addition, the comments people receive show no such rise at all. Why, then, do published studies disagree? Reddit community types were already drifting apart before ChatGPT existed, at rates comparable to every published estimate, and without same-time controls, that drift can look like an effect of generative AI. Our own largest estimate, an 18\% fall in posts to low-stakes curiosity communities, matches its pre-existing trend. Humans still ask humans for help, and, as far as detection can tell, humans still answer them.
\end{abstract}

\section{Introduction}

For as long as the web has existed, people have used it to ask strangers for help with a diagnosis, a lease, a tax form, or a broken build. Generative AI is the first technology that can plausibly answer those questions directly. Whether people still ask strangers for help is therefore an open question about the future of the social web. ChatGPT was released on 30 November 2022. Within two years, several studies have asked whether it has reduced the volume of questions people ask other people online, and these studies cannot agree on an answer. \citet{delrio2024} report a 25\% fall in Stack Overflow activity against a comparison set of other websites. \citet{xue2026} report an average 14\% decline on the same platform, up to 27.9\% over time (the earlier conference version of the same study reported 2.6\%; \citealp{xue2023}). \citet{gaohahn2025} study 17 Reddit advice communities and report a decline of 0.227 posts per 10,000 members per day, which is 8.3\% of their reported mean. \citet{burtch2024} find no decline in Reddit developer communities. \citet{koonchanok2026} finds no decline in a test-prep community and a rise in one category of help-seeking. These estimates differ by a factor of ten and do not agree on direction. One explanation is that different platforms and communities responded differently, and that may be part of it. We show that a second explanation is enough on its own. None of the five studies compares against a group of communities that AI cannot easily substitute for, over the same months, and the communities they compare against were already moving apart before ChatGPT existed.

Our design supplies this missing comparison. We compare informational communities to hobby communities chosen to cover their size range, on the same platform in the same months, so that anything that moved Reddit as a whole affects both groups equally. We also rerun the full analysis at 66 earlier dates before ChatGPT's release, which we call placebo dates. No ChatGPT effect can exist at these dates, meaning that whatever the analysis reports there reflects pre-existing trends between the community groups rather than any effect of AI. Two results emerge from this design. First, informational help-seeking did not fall. We estimate a 5.1\% increase in posts, and our results rule out any decline larger than 3.4\%, which excludes every decline published in the literature except the smallest. Second, at placebo dates when no ChatGPT effect exists, our results find effects as large as the published ones. In other words, community types were already drifting apart at rates comparable to every published estimate. A design without same-time controls could therefore have obtained any of them, in either direction, from drift alone.

Changes in post counts say nothing about who wrote the posts themselves. It is possible that humans left and machine-generated posts filled the gap, leaving post count totals steady while human posting fell. We test this possibility directly by scoring 274{,}411 sampled posts with an AI-text detector, namely, Fast-DetectGPT \citep{bao2024}. The detector's average score goes through the same comparison as the counts, in which any fixed bias in the detector cancels out. We find that AI-written posts rose 2--3 percentage points more in informational communities than in hobby communities, depending on which of four LLMs we use to calibrate the detector. This rise is not large enough to hide the smallest decline of 8.3\% previously reported in the literature for Reddit. While the questions may still be human, the answers, in the form of the comments under a given post, could be machine-generated. We repeat the same test on 223{,}775 comments, adding a second detector, Binoculars \citep{hans2024}, because Fast-DetectGPT could not reliably separate machine-generated comments from human ones. Neither detector finds extra AI content accumulating in informational communities' answers.

One of our own estimates shows how ordinary platform drift can look like an effect of generative AI. In low-stakes curiosity communities, the places where an AI chatbot could most plausibly replace a human, post volume falls 18.3\% relative to controls after the launch, with $p = 0.002$. Read on its own, this could be interpreted as people taking their questions to AI instead. However, these communities were already shrinking at roughly 18\% a year against the same controls before ChatGPT existed, and that pre-existing trend accounts for the estimate in full.

This study makes four contributions. (1)~The first estimate of ChatGPT's effect on public help-seeking volume against contemporaneous control communities that AI cannot easily substitute for, ruling out any decline in posting larger than 3.4\%. (2)~A detector-based test of whether AI-written text is inflating post counts, calibrated on the platform's real posts against four LLMs (three widely used during the study window, and one later model as a conservative stress test), for both questions and answers. (3)~An explanation for why the published record disagrees, in the form of measured pre-existing trends as large as every reported effect. (4)~A general diagnostic for studies of platform-scale events, which reruns the full analysis at placebo dates before the event and measures the false effects the design produces on its own.

\section{Related Work}

\paragraph{Help-seeking in online communities.} Asking strangers for help is one of the oldest behaviors on the social web, and one of the most extensively studied. Question-answering communities produced answers fast and well enough that they became the standard resource in their domains \citep{anderson2012}. The support exchanged in them comes in two kinds: informational support passes along facts and procedures, and emotional support gives recognition \citep{cutrona1992}. The two kinds should plausibly respond differently when a machine can supply them. Informational support is valued for the content of the reply, and a system that produces correct facts and procedures on demand is therefore a direct substitute. Emotional support, on the other hand, is valued for who it comes from. Recognition counts when a person chooses to give it, and people rate the same supportive message as less helpful once they know it came from an AI \citep{yin2024}. Reddit hosts both kinds of support communities, with communities organized around practical questions operating alongside communities where anonymity allows disclosures that members cannot make elsewhere \citep{dechoudhury2014}.

\paragraph{Estimates of ChatGPT's effect on help-seeking.} Table~\ref{tab:record} lists the five published estimates we compare our results against, along with the comparison group each uses. \citet{delrio2024} compare Stack Overflow to other websites, which leaves anything that affected only Stack Overflow inside the estimate. \citet{gaohahn2025} and \citet{burtch2024} compare the same communities to themselves one year earlier, which means any long-running trend is not controlled for. \citet{koonchanok2026} uses an interrupted time series on a single community with no separate control group, stating that assumption plainly. Beyond help-seeking behavior, the literature has also examined ChatGPT's effect on contribution volume. \citet{shanqiu2025} report reduced voluntary contribution on Stack Overflow, and \citet{lyu2025} report bounded effects on Wikipedia contribution. Closest to our setting, \citet{zhang2026} estimate the effect of AI answers in Google search on Reddit itself, finding that Google's AI Overviews raised comments in indexed communities by 12\% with no statistically detectable movement in posts, an estimate our 2024--25 window relies on. The timing of the comparison also shapes what it can find, because adoption was fast but uneven. Nearly 40\% of the US population aged 18--64 was using generative AI within two years of the launch \citep{bick2024}, skewed toward younger workers. A design that measures the launch's first six months catches the technology early in its adoption, which is why we also report a 2024--25 window.

\begin{table}[t]
\centering\footnotesize
\setlength{\tabcolsep}{3.5pt}
\begin{tabular}{@{}llc@{}}
\toprule
Study & Comparison group & Estimate \\
\midrule
del Rio-Chanona et al.\ 2024 & other websites & $-25\%$ \\
Xue et al.\ 2026 & within Stack Overflow & $-14\%$ \\
Gao and Hahn 2025 & same subs, prior year & $-8.3\%$ \\
Burtch et al.\ 2024 & same subs, prior year & none \\
Koonchanok 2026 & none (ITS) & none / $+$ \\
\midrule
This paper & low-exposure, contemp. & $+5.1\%$ \\
\bottomrule
\end{tabular}
\caption{Published estimates at ChatGPT's launch. No prior study compares against a contemporaneous group of communities that AI cannot easily substitute for. \citet{gaohahn2025}'s coefficient is in posts per 10,000 members per day, and we convert it against their reported mean of 2.742.}
\label{tab:record}
\end{table}

\paragraph{Mechanism.} Two plausible explanations make opposite predictions about which communities would lose volume after ChatGPT's release. The social-fabric explanation, stated by \citet{burtch2024}, holds that people come to peer communities for recognition and shared experience, and that communities whose value is social should therefore be protected. The privacy explanation, reported by \citet{gaohahn2025} from \citet{wester2024}, holds that a machine that does not judge is easier to confide in, and that the most personal communities should therefore decline first. Deciding between them requires comparing community types directly to each other, and Section~5 shows that this across-type comparison is the one pre-existing drift distorts most. We therefore estimate each community type against controls, the comparison our placebo dates show to be reliable, and leave the across-type question as one the data around ChatGPT's launch cannot yet answer cleanly. Settling the two explanations would also require account-level posting histories, which we do not collect because our sample includes grief, abuse-survivor, and crisis communities, where following individual accounts would pose serious re-identification risks.

\paragraph{Machine-generated-text detection.} The second half of this study uses statistical AI-text detectors that need no training data, a line of work running from probability-curvature tests \citep{mitchell2023} through their efficient single-pass variant \citep{bao2024} to paired-model perplexity ratios \citep{hans2024}. These methods score a document (i.e., a body of text) by how surprising its words are to a language model, which makes them reproducible and independent of any training set. Detector scores can be wrong for any single document, and they are documented to be biased against non-native English writers \citep{liang2023}. Our design avoids both problems. No document is ever labeled AI or human, and the quantity we estimate is the change in a group's average score relative to controls, in which any detector bias that stays constant over time cancels out. A change in who is writing within a group over time would not cancel, and we discuss this in the Limitations section.

\section{Data and Sample}

\subsection{Monthly post counts}

We use monthly post counts from the Arctic Shift Reddit archive \citep[a maintained successor to the Pushshift collection;][]{baumgartner2020}, which serves per-community monthly counts through a public endpoint. Our analyses of post count change uses these counts and nothing else; no post text or author information is read.

We began from 157 candidate communities, assigned to four groups before any outcome was computed. The rule follows the support-type distinction \citep{cutrona1992} and asks what a satisfying reply would look like. Group~I is informational advice, where a language model can supply the answer directly because the reply is a fact, a procedure, or a diagnosis (r/legaladvice, r/personalfinance, r/techsupport, r/AskDocs and 22 others). Group~H is human-anchored support, where the asker wants a witness rather than a fact, and the reply's value depends on coming from a person (r/GriefSupport, r/relationship\_advice, r/SuicideWatch and 26 others). Group~Q is low-stakes curiosity, where the question is asked for the fun of the answer (r/explainlikeimfive, r/tipofmytongue, r/whatisthisthing and three others). Group~C is the control group, hobby communities with substantially less plausible exposure to substitution (r/knitting, r/houseplants and 88 others), chosen to cover the treated groups' size range; their posts are largely projects, photos, and community talk rather than questions a model could answer. The full assignment is in Appendix~C.

A community stays in our sample if it passes four mechanical filters on pre-ChatGPT data. It must have been founded on or before December 2017, so that the full 59 months of pre-ChatGPT history used by the placebo dates exist for every community. It must have a median of at least 100 posts a month, and at least 50 posts in 90\% of months; these floors keep the analysis away from small communities where a handful of posts can move the estimate. Finally, it must have no missing months in that history, so that the same communities enter every comparison. Six communities failed these filters: r/AskALawyer and r/DadForAMinute due to their founding date, r/whittling for not meeting the volume floor, and r/AskDentists, r/FirstTimeHomeBuyer, and r/crossword for not meeting the 50-post requirement. The surviving sample is 151 communities, with 26 in I, 29 in H, 6 in Q, and 90 in C.

\subsection{Text samples for the detector analysis}

The detector analysis (Section~\ref{sec:detector}) additionally samples document text through the Arctic Shift archive. It draws up to 100 posts and 100 comments per community per month over two windows, January 2021 to November 2022 and January 2024 to November 2025. The first window predates ChatGPT and is therefore a human-written baseline; the second is chosen a year after ChatGPT's release to give machine-generated content the most time to accumulate. The cap of 100 gives every community-month equal weight in the scored sample. Our confidence intervals come from resampling communities, and differences between communities dominate the uncertainty, meaning that scoring more documents per month would not make the estimate more precise. This process yields 680,267 posts and 689,468 comments. No author field is ever requested, deleted and removed items are dropped, and scoring retains only the group, community, month, length, and score of each document; no text, title, or identifier survives past the scoring step. Documents shorter than 200 characters are excluded because they are too short for the detector to score reliably. The exclusion means the scored documents are a subset of the counted ones, and if the share of documents passing the length floor moved differently across groups between periods, the exclusion itself could produce a score shift. When that this was the case for posts but not for comments, and Section~\ref{sec:detector} measures the shift and adjusts for it; Appendix~D reports the shares by group and period.

\section{Design}

\subsection{The two-way fixed-effects specification}

We ask whether each treated group's volume moved against the controls at the launch, controlling for fixed differences between communities and for platform-wide events in any given month. We apply a two-way fixed-effects regression to monthly counts as:

\begin{equation}
\log(1 + y_{st}) = \alpha_s + \gamma_t + \sum_{g \in \{I,H,Q\}} \delta_g \, \mathbf{1}[s \in g]\cdot \mathrm{Post}_t + \varepsilon_{st},
\end{equation}

where $y_{st}$ is the number of posts community $s$ received in month $t$, with community fixed effects $\alpha_s$, month fixed effects $\gamma_t$, and group C as the omitted reference. Each $\delta_g$ is a difference-in-differences: the change in group $g$'s monthly posts before versus after the launch, minus the same change in the control communities. The coefficients are in log points, which read approximately as percentage changes (a coefficient of $+0.05$ is close to a 5\% increase), and we convert them to exact percentages for simplicity. Standard errors are clustered by community, since each community's errors are correlated over time \citep{bertrand2004}, and we report wild cluster bootstrap-S $p$-values alongside \citep{cameron2008,mackinnonwebb2018}. The design's key assumption is that the groups were not already trending apart before the launch, and we test this directly at the placebo dates described in Section~\ref{sec:placebo}.

Our main quantity of interest is $\delta_I$, the movement of informational communities relative to controls after ChatGPT's release. It asks whether the communities a language model can answer directly lost volume. $\delta_H$ is the same movement for human-anchored support communities, and the difference $\Delta = \delta_I - \delta_H$ would say which type of community lost more volume, the question the two explanations of Section~2 disagree on. However, informational and human-anchored communities were already drifting apart before ChatGPT existed, at rates close to the effects under study (Section~\ref{sec:placebo}). As a result, a $\Delta$ estimated at the real launch cannot be told apart from that drift. Correcting for the drift does not help; the corrected estimators, including one that asks whether the drift accelerated at the launch, are either still biased or too noisy to detect anything (Appendix~B). We therefore report $\Delta$ without interpreting it.

\subsection{The six-month comparison window}

Reddit informational communities have strong within-year swings, and a window that compares different calendar months would fold those seasonal swings into the estimate. We measure this directly: relative to their annual means, group~I averages $-0.008$ log points over December to May and group~H $-0.047$, meaning an unmatched window would bias the comparison between the two groups by about 4\%. The primary window therefore compares six pre-ChatGPT months (December 2021 to May 2022) against the same six calendar months one year later (December 2022 to May 2023), with June to November 2022 dropped. Comparing the same calendar months removes the seasonal problem entirely.

Two factors set the window length. First, the spread of community-level changes grows as the two windows move apart, from 0.310 at a three-month gap to 0.349 at 24 months. If the noise were month-to-month, that spread would shrink instead. The spread is therefore slow drift at the community level, meaning a nearby window is both more precise and less exposed to drift. Second, a one-year gap matches the year-over-year comparison \citet{gaohahn2025} use, which makes the two estimates directly comparable.

We keep the full 59 months of pre-ChatGPT history for the placebo dates, and report an estimate over that full history as a robustness check. Section~\ref{sec:longwindow} reports a second, longer window (2021--22 against 2024--25) that measures the effect two to three years after the launch rather than in its first six months. The primary window is early in ChatGPT's adoption; the long window is late, but overlaps with other events that affected Reddit, namely Google's licensing deal with Reddit (February 2024) and the rollout of Google's AI Overviews (May 2024). We therefore report the results from both window comparisons.

\subsection{What the design can detect}

A null result is only meaningful if the design could have detected a real decline, and we therefore quantify what it can detect before reporting any estimate. The dominant source of noise is not month-to-month variation but slow drift at the community level. After seasonal matching, the community-level spread of changes at the primary window is 0.194. With 26 treated communities, the smallest decline the design can reliably detect (at 80\% power and a 5\% significance level) is about 14\%. Most of the precision comes from two choices. Matching the calendar months removes a bias of about 4\%, and keeping the two windows close together limits drift, which grows with distance. Adding more months to the window would therefore not make the design more sensitive.

\subsection{Placebo launch dates before ChatGPT}
\label{sec:placebo}

If, before ChatGPT, treated and control communities were growing or shrinking at the same rate, then a difference-in-differences estimated at any date before its release should return zero. We therefore rerun the estimate at every earlier date that allows the same comparison. Each placebo date is treated as a pretend launch: the six months before it serve as the pre-period, the same six calendar months one year later serve as the post-period, and the months in between are dropped, exactly as in the primary comparison. There are 66 such dates, the earliest starting in January 2017 and the last ending before ChatGPT's release. Since nothing launched at these dates, any nonzero estimate is the pre-existing trend between the groups, the same trend that would contaminate an estimate at the real launch date. Appendix~A reports all 66 placebo date estimates.

\subsection{The detector difference-in-differences}
\label{sec:detectordesign}

Our final analysis asks whether posts in our communities of interest are increasingly LLM-written in the post-ChatGPT era. We use the same design as before, but with a different outcome variable, namely, the mean Fast-DetectGPT score of a community's sampled documents in a period. For a scoring model $p_\theta$ and document $x_1 \ldots x_n$, the score is the conditional probability curvature

\begin{equation}
d(x) = \frac{\sum_t \log p_\theta(x_{t+1} \mid x_{\le t}) - \sum_t \mathbb{E}_{v}\,[\log p_\theta(v \mid x_{\le t})]}{\sqrt{\sum_t \mathrm{Var}_{v}\,[\log p_\theta(v \mid x_{\le t})]}},
\end{equation}

which is higher for machine-generated text \citep{bao2024}. We use GPT-Neo-1.3B \citep{gptneo2021} as the scoring model, truncate documents at 512 tokens, and use the score itself as the outcome rather than labeling any document as AI or human. The quantity we estimate is $(\bar{d}{g,\mathrm{post}} - \bar{d}{g,\mathrm{pre}}) - (\bar{d}{C,\mathrm{post}} - \bar{d}{C,\mathrm{pre}})$, the same difference-in-differences as before with the community as the unit. Communities are equally weighted, and confidence intervals come from resampling communities within groups. The pre-period (January 2021 to November 2022) serves as the pre-ChatGPT baseline. A detector that over-scores some groups of writers \citep{liang2023} adds the same offset to both periods and both groups, and cancels. The comparison therefore measures how much more machine-like a group's text became relative to controls.

To turn score shifts into shares of documents, we measure how the detector reacts to known machine-generated text of this exact kind. Specifically, we sample 300 real pre-ChatGPT posts from groups I and H, and have four LLMs each produce a matching document, scored by the identical pipeline. Three of them (GPT-4o, Claude Sonnet 4, and DeepSeek-V3) were widely used during the post window; the fourth (GLM-5.2) is a 2026 release included as a conservative stress test. For posts, the LLM writes a post body from the real title; for comments, it writes a reply to the real post's title and body, the way a person pasting a question into a chatbot would obtain an answer. If generated documents score $S$ units above the human baseline, a value we call the separation, then a shift of $\delta$ in the mean score corresponds to a change of $\delta/S$ in the share of machine-written documents. The same calibration yields an AUROC, a standard measure of whether the detector can distinguish a given LLM's text from human text at all. Appendix~E reports prompts, success rates, and the calibration procedure.

\section{Results}

\begin{figure*}[t]
\centering
\includegraphics[width=0.95\textwidth]{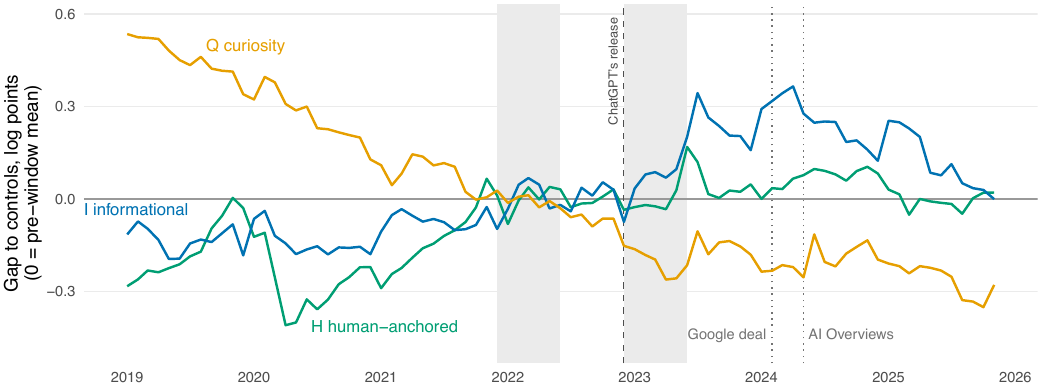}
\caption{Each line is a group's mean $\log(1+y)$ gap to the controls, for informational (I, blue), human-anchored (H, green), and curiosity (Q, orange) communities, normalized to its mean over the primary pre-period; shaded bands mark the pre and post windows of the primary comparison, and the dashed line marks ChatGPT's release. Curiosity communities were falling steeply for years before ChatGPT. Informational communities show no sustained movement at the launch and rise in 2024--25, a period that also contains Google's licensing deal with Reddit and the rollout of AI Overviews (dotted lines).}
\label{fig:eventstudy}
\end{figure*}

\subsection{Informational help-seeking did not decline}

If people began taking their questions to ChatGPT instead of to each other, a dip in posts should appear first in the communities whose questions a model can answer directly. Group I's volume, in other words, should fall against controls at the launch. Table~\ref{tab:main} instead reports $\hat{\delta}_I = +0.050$, a 5.1\% increase in monthly posts, with a 95\% confidence interval of $[-0.034, +0.135]$ and $p = 0.24$ (bootstrap $p = 0.28$). The interval excludes any decline larger than 3.4\%. This suggests that informational help-seeking did not decline after ChatGPT's launch.

Excluding declines beyond 3.4\% places our estimate below every published decline except one. Table~\ref{tab:bench} compares each against our interval. The interval rules out a decline the size of Stack Overflow's 25\% \citep{delrio2024}, the 18.2\% reported for casual users \citep{quinngutt2025}, the 8.3\% reported for Reddit advice communities \citep{gaohahn2025}, and the 14\% average decline of \citet{xue2026}. The one published number it cannot rule out is the 2.6\% from that study's earlier conference version \citep{xue2023}, which is smaller than the smallest decline our design can reliably detect.

\begin{table}[t]
\centering\footnotesize
\setlength{\tabcolsep}{3.5pt}
\begin{tabular}{@{}lrrrr@{}}
\toprule
 & $\hat{\delta}$ & SE & 95\% CI & $p$ \\
\midrule
$\delta_I$ informational & $+0.050$ & 0.043 & $[-0.034, +0.135]$ & 0.24 \\
$\delta_H$ human-anchored & $-0.018$ & 0.033 & $[-0.083, +0.046]$ & 0.57 \\
$\delta_Q$ curiosity & $-0.202$ & 0.063 & $[-0.327, -0.078]$ & 0.002 \\
$\Delta = \delta_I - \delta_H$ & $+0.069$ & 0.047 & $[-0.025, +0.162]$ & 0.15 \\
\bottomrule
\end{tabular}
\caption{Difference-in-differences against control communities, primary window. Coefficients are log points of monthly posts relative to controls; a coefficient of $-0.202$ corresponds to an 18.3\% decline in posts. Wild cluster bootstrap-S $p$-values are 0.28, 0.56, 0.01, and 0.16.}
\label{tab:main}
\end{table}

\begin{table}[t]
\centering\footnotesize
\setlength{\tabcolsep}{2pt}
\begin{tabular}{@{}lrc@{}}
\toprule
Benchmark & log pts & Excluded? \\
\midrule
Stack Overflow $-25\%$ (del Rio-Chanona) & 0.288 & yes \\
Casual users $-18.2\%$ (Quinn and Gutt) & 0.201 & yes \\
Stack Overflow $-14\%$ (Xue et al., ISR) & 0.151 & yes \\
Reddit advice $-8.3\%$ (Gao and Hahn) & 0.086 & yes \\
Stack Overflow $-2.6\%$ (Xue et al., conf.) & 0.026 & no \\
\bottomrule
\end{tabular}
\caption{Published declines compared against our 95\% confidence interval. Every decline except the smallest falls outside the interval and is ruled out for informational help-seeking on Reddit.}
\label{tab:bench}
\end{table}

Figure~\ref{fig:eventstudy} shows each group's posting trend over time. Group Q declines throughout the whole period, from about $+0.5$ in 2019 to about $-0.3$ by late 2025, with no visible change in that slide at the launch. Group H starts low, falls further in 2020, climbs back to zero by late 2021, and stays near zero from the launch onward. Group I, the group the substitution hypothesis concerns, sits near zero through the pre window, dips briefly at the release, and climbs to about $+0.1$ by the end of the post window; averaged over the six months the difference-in-differences compares, the post window sits 0.05 log points above the pre window, the estimate of Table~\ref{tab:main}. Group I's large movements all come after the primary windows close. Its gap spikes to $+0.34$ in June--August 2023, when many communities restricted posting in protest of Reddit's API pricing changes, falls back, rises again to $+0.36$ in early 2024, after Google's licensing deal and before the rollout of AI Overviews, and then drifts back toward zero through 2025. The protest months fall in the gap between the two windows and enter no estimate; the 2024--25 movements average to the long-window estimate that Section~\ref{sec:longwindow} examines.

\subsection{Every estimate matches its pre-existing trend}

\begin{figure}[t]
\centering
\includegraphics[width=0.9\columnwidth]{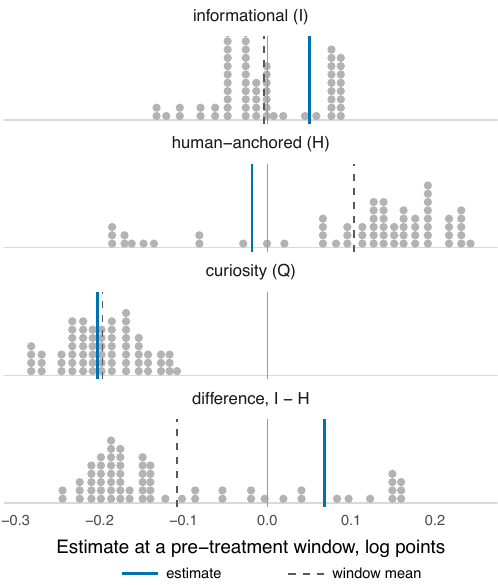}
\caption{Each panel shows the estimate at all 66 placebo dates (dots; dashed line their mean) beside the estimate at the real launch (solid line). $\delta_I$'s placebo estimates are centered on zero and contain its launch estimate. $\delta_H$'s are centered far from zero. Every one of $\delta_Q$'s 66 placebo estimates is negative, and its launch estimate falls among them. The difference $\Delta$ inherits the drift of its parts, and an estimate of it at the launch would be uninterpretable.}
\label{fig:sweep}
\end{figure}

If the launch moved any group, its estimate should stand apart from the estimates the same regression returns at the placebo dates. We find that it does not. The movement each group shows after ChatGPT's release is the movement the same regression reports at dates where no effect exists. Table~\ref{tab:placebo} and Figure~\ref{fig:sweep} place each launch estimate beside the 66 placebo estimates.

\begin{table}[t]
\centering\small
\begin{tabular}{@{}lrrr@{}}
\toprule
 & Launch & Placebo mean & Placebo range \\
\midrule
$\delta_I$ & $+0.050$ & $-0.004$ & $[-0.133, +0.093]$ \\
$\delta_H$ & $-0.018$ & $+0.104$ & $[-0.187, +0.242]$ \\
$\delta_Q$ & $-0.202$ & $-0.197$ & $[-0.285, -0.108]$ \\
$\Delta$ & $+0.069$ & $-0.108$ & $[-0.245, +0.162]$ \\
\bottomrule
\end{tabular}
\caption{Estimates at the real launch beside the mean and range of the 66 placebo dates, all in log points of monthly posts relative to controls. Since ChatGPT did not exist at the placebo dates, the true effect at each is zero.}
\label{tab:placebo}
\end{table}

Because neighboring windows share most of their months, the 66 estimates are correlated rather than independent. This means their spread cannot be used as a standard error, and we use it only descriptively. The overlap, however, does not bias their mean, and the mean is the only quantity our estimate rests on. The drifts these placebos reveal are large. Measured in monthly posts against size-comparable controls, human-anchored communities gain 0.104 log points a year, and curiosity communities lose 0.197, which puts the two group types about 0.30 log points apart after a single year with no event at all. Figure~\ref{fig:eventstudy} shows the same split in the raw series. Four in five of the windows return a negative $\Delta$, and the window immediately before the launch gives $-0.142$. Pre-existing drift is therefore as large as the effects being estimated at ChatGPT's launch.

\subsection{The curiosity-community decline matches its pre-treatment trend}
\label{sec:deltaq}

If language models were substituting for the answers people come to these communities for, curiosity communities should be the first to see their post counts fall, since a model answers exactly the questions r/explainlikeimfive exists to answer. The data largely appear to agree. The largest coefficient we estimate is $\delta_Q = -0.202$ log points, an 18.3\% decline in monthly posts to curiosity communities, at $p = 0.002$. Its placebo mean, however, is $-0.197$. In other words, these communities were losing 18\% a year against size-comparable controls before ChatGPT was released, and the launch estimate differs from that trend by only 0.006 log points, far below the 0.090 the design would need to detect a change.

The decline is also much older than ChatGPT. Curiosity communities were losing posts at 18\% a year for at least five years before it existed (Figure~\ref{fig:eventstudy}). Whatever was absorbing low-stakes curiosity questions, whether better search, instant answer boxes, or the platform's own aging, had been doing so well before the launch, and the launch produced no detectable change in the rate.

\subsection{The estimate survives sample and window changes}

To test whether the primary estimate depends on which communities enter it or on the choice of pre-window, we refit it on subsamples of the main sample and on the 59-month pre-window. Table~\ref{tab:robust} reports the four variants. $\hat{\delta}_I$ is $+0.084$ log points on the 60\% of communities held out from all exploratory work and $-0.003$ on the 40\% used for exploration, and both intervals overlap the primary estimate. Dropping any one community and refitting always lands in $[+0.035, +0.071]$ with the primary's sign, and no single community drives the estimate. Using the 59-month pre-window instead of the matched six months gives $+0.136$, further from zero in the same direction, consistent with the longer window absorbing more of the upward drift.

\begin{table}[t]
\centering\small
\begin{tabular}{@{}lrr@{}}
\toprule
Cell & $\hat{\delta}_I$ & SE \\
\midrule
Primary & $+0.050$ & 0.043 \\
Held-out 60\% of the sample & $+0.084$ & 0.051 \\
Exploratory 40\% of the sample & $-0.003$ & 0.075 \\
59-month pre-window & $+0.136$ & 0.071 \\
Leave-one-out (26 refits) & $+0.035 \ldots +0.071$ & --- \\
\bottomrule
\end{tabular}
\caption{Robustness of the primary estimate across subsamples and the alternative pre-period, in log points of monthly posts relative to controls. Every cell except the near-zero exploratory subsample shares the primary's sign, and none comes close to the $-0.086$ log points that an 8.3\% decline would require.}
\label{tab:robust}
\end{table}

\section{The 2024--25 Window}
\label{sec:longwindow}

The primary window ends six months after the launch, before ChatGPT adoption reached scale. To test whether our results depend on this choice, we run the same design over a longer window, with a pre-period of January 2021 to November 2022, a post-period of January 2024 to November 2025, and the intervening months (December 2022 to December 2023) excluded, so that the post-period follows the years of early adoption rather than containing them. The sample with complete data over this longer period is 149 communities (24~I, 29~H, 6~Q, 90~C).

A decline that required adoption to reach scale should appear in this window. However, the estimate instead runs the other way, with $\hat{\delta}_I = +0.218$ log points ($p = 0.04$), a 24\% \emph{rise} in monthly posts to informational communities against controls, and $+0.126$ after subtracting the one available placebo run at this window length.

We do not read this result as an effect of ChatGPT, for two reasons. First, two changes in how Reddit reaches readers fall inside the post-period, namely, Google's licensing deal with Reddit (February 2024) and the rollout of AI Overviews (May 2024), and Reddit content is disproportionately valuable to search engines \citep{vincent2019}. Their net effect on posting is unsettled, because summarizing an answer on the results page removes a reason to visit Reddit, while surfacing Reddit threads more prominently sends more visitors to them. The one direct estimate finds that AI Overviews raised daily comments in indexed communities by 12\%, with no statistically detectable movement in posts \citep{zhang2026}. Second, the post-period contains whatever share of posts is now machine-generated. Post counts cannot separate an effect of ChatGPT from these overlapping changes, which land on exactly the informational questions group~I answers (Figure~\ref{fig:eventstudy} marks both search events). Only one placebo window exists at this length, and the drift adjustment therefore carries no interval. The full table, including the placebo run in which $\delta_Q$ reaches $-0.561$, is in Appendix~F.

The window still puts a useful number on how large any hidden decline could be. By the 2024--25 post-period, generative-AI adoption had reached roughly 40\% of the US working-age population \citep{bick2024}, widespread enough for substantial substitution to be plausible. For the observed $+24\%$ to conceal a decline the size of the smallest published Reddit estimate, 8.3\%, the post-period would have to carry about a third more posts than it otherwise would, relative to controls, from machine writing and search-driven arrivals combined. Concealing the largest published estimate, 25\%, would take about two thirds more. The detector analysis of Section~\ref{sec:detector} runs on this same window and puts the differential increase in machine-like posting relative to controls at roughly 2--3 percentage points, and the one direct estimate of the AI Overviews effect reports no movement in posts \citep{zhang2026}. The long window therefore cannot deliver a clean estimate, but it makes a large late-arriving decline difficult to reconcile with the observed data.

\section{Testing Whether Machine-Generated Content Masks a Decline}
\label{sec:detector}

If machine-generated posts piled up in informational communities faster than in controls, the counts could hold steady while human participation fell, such that $\delta_I^{\mathrm{observed}} = \delta_I^{\mathrm{human}} + \text{inflation}$. For this inflation to make our confidence interval stop ruling out the 8.3\% decline, machine-written posts would have to account for about 5.1 percentage points of informational-community posts, relative to controls. The detector windows are those of Section~\ref{sec:longwindow}, comparing 2021--22 with 2024--25. Machine-generated content should be far more prevalent there than in early 2023, so the analysis measures accumulation where it should be largest; for the primary window it serves as a later-period consistency check.

\subsection{Which documents can be scored}

Fast-DetectGPT, the detector specified in Section~\ref{sec:detectordesign} and calibrated in Appendix~\ref{app:calib}, scores only documents of at least 200 characters. Before reading any score, we therefore check whether that filter selects differently across groups and periods. On the comment side it does not. The share of documents passing that floor is flat before and after ChatGPT's release in every group (group~I moves from 0.472 to 0.436, controls from 0.267 to 0.272; Appendix~D), and every control community qualifies in both periods. On the post side, however, it does. The share of sampled posts carrying at least 200 characters of body text rises in every group between the periods, from 0.54 to 0.75 in group~I and from 0.21 to 0.42 in controls (Appendix~\ref{app:textshare}).

The posts we can score are therefore a larger and differently chosen share of the posts we count in the later period, and the change is bigger in the control communities. Every scored post comes with its length, which lets us compare posts of similar length instead. Giving each length group the weight it had in the earlier period leaves the result unchanged, $+0.055$ ($[+0.013, +0.099]$) against $+0.053$ without the adjustment (Appendix~\ref{app:calib}). Posts too short to score are missing from both periods, and no adjustment can recover them. Ten of ninety control communities have no posts long enough to score in one period and drop from this part of the analysis.

\begin{figure}[t]
\centering
\includegraphics[width=0.9\columnwidth]{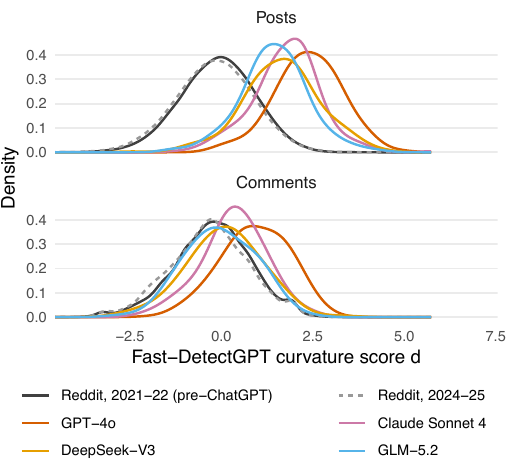}
\caption{The detector separates machine-generated text from human text for posts, only partially for comments, and the distribution of real Reddit documents barely moves between 2021--22 and 2024--25. Densities of Fast-DetectGPT scores for real documents in the two periods (grey; the 2021--22 period is the pre-ChatGPT baseline) and matched documents from the four calibration LLMs (color), for posts (top) and comments (bottom). On posts every LLM's curve sits clear of the bulk of the human distribution (AUROC 0.875--0.966); on comments only GPT-4o separates (0.794) while the other three LLMs' body-conditioned replies nearly coincide with the human curve (0.552--0.664).}
\label{fig:detector}
\end{figure}

To establish what the detector can see, we calibrate it against the machine-generated documents from the four LLMs described in Section~\ref{sec:detectordesign}. Figure~\ref{fig:detector} shows the score distributions. Fast-DetectGPT reliably identifies posts from all four LLMs, with AUROCs ranging from 0.966 (GPT-4o) to 0.875 (GLM-5.2) and separations of $+2.48$ to $+1.56$ score units above the human baseline. However, it does not reliably detect machine-generated comments. GPT-4o replies remain detectable (AUROC 0.794, separation $+1.15$), while replies from the other three LLMs sit almost on top of the human distribution, with AUROCs of 0.664, 0.585 and 0.552 and separations of $+0.57$, $+0.32$ and $+0.20$. Because that leaves most modern models invisible on comments, we score the comments again with a second detector, Binoculars \citep{hans2024}, which uses two language models instead of one and is more accurate on short text. Binoculars can detect comments from GPT-4o (AUROC 0.939) and Claude Sonnet 4 (0.850), still cannot see DeepSeek-V3 or GLM-5.2 (0.768, 0.698), and agrees with Fast-DetectGPT on posts for all four LLMs (0.896--0.967). Appendix~\ref{app:calib} reports both calibrations in full. Lastly, Figure~\ref{fig:detector} also shows that posts and comments published in 2024--25 score almost identically to those from 2021--22, with no visible movement toward the range where machine-generated text sits.

\subsection{Posts show differential inflation of 2--3\%}

This estimate never labels any individual document as machine-generated. A detector used with a cutoff would wrongly flag human writers, at a rate that varies with dialect and fluency \citep{liang2023}. We average a continuous score instead, which needs no cutoff, and Ithe differencing removes whatever score level the detector attaches to a community's writers. The estimate is also relative rather than absolute. We cannot say what share of 2024--25 posts is machine-generated, since that would need our calibration text to stand in for all machine-generated text on Reddit. We can only estimate whether machine-generated text piled up faster in one group than in another.

\begin{figure}[t]
\centering
\includegraphics[width=0.9\columnwidth]{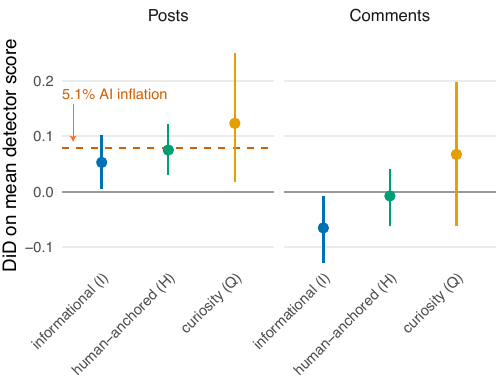}
\caption{The rise in machine-written posts stays below the 5.1\% threshold at which the volume interval would no longer rule out the 8.3\% decline, and informational communities' comments show no rise at all (the comment points shown are Fast-DetectGPT's; the second detector bounds the comment estimate to a tight null, Section~7.3). Points are difference-in-differences on mean detector score, with 95\% bootstrap intervals over communities. The dashed line is the score shift that a 5.1\% share of AI-written posts would produce under the strictest of the four calibrations (the GLM-5.2 separation); the $\delta_I$ point stays under it, though its interval crosses.}
\label{fig:detectordid}
\end{figure}

Machine-written posts propping up informational community counts would raise these communities' mean score against controls. Table~\ref{tab:detector} and Figure~\ref{fig:detectordid} report the difference-in-differences on mean score over 274{,}411 scored posts from the 139 communities with scoreable text in both periods, with each community contributing one mean per period. Every group's mean score fell between the periods (I $-0.102$, H $-0.079$, Q $-0.031$, C $-0.155$ score units), a platform-wide drift that the comparison with controls removes. The shift for informational communities relative to controls is $+0.053$ score units (95\% CI $[+0.005, +0.102]$), and it arises because control communities fell furthest rather than because informational communities rose.

Dividing the $+0.053$ score-unit shift by each LLM's measured separation converts it into an extra machine share of posts, from $+2.1\%$ (CI $[0.2\%, 4.1\%]$) under the GPT-4o calibration, through $+2.8\%$ and $+3.0\%$ under Claude Sonnet 4 and DeepSeek-V3, to $+3.4\%$ (CI $[0.3\%, 6.5\%]$) under GLM-5.2. Every point estimate sits under the 5.1\% derived at the opening of this section, the share at which the volume interval would no longer rule out an 8.3\% decline. Only the strongest calibration's interval also rules out that threshold; under the three weaker separations the upper bound reaches 5.4\% to 6.5\% (Appendix~E tabulates the conversions). The dashed line in Figure~\ref{fig:detectordid} draws the threshold under the GLM-5.2 separation, the calibration that places the bar closest to the estimate, and there the $\delta_I$ interval crosses the line while the point estimate stays below it. Three of the LLMs were widely used during the post window, and GLM-5.2, a more recent model, is included as a conservative stress test; together the four keep the conversion from leaning on any single model's detectability. Even when calibrated on the least detectable model we measured, the central estimate stays below the threshold at which the 8.3\% decline would no longer be ruled out.

\begin{table}[t]
\centering\footnotesize
\setlength{\tabcolsep}{3.5pt}
\begin{tabular}{@{}llrrr@{}}
\toprule
Arm & & $\hat{\delta}^{d}$ & 95\% CI & implied share \\
 & & \multicolumn{2}{c}{\footnotesize score units} & \footnotesize of documents \\
\midrule
Posts & $\delta_I$ & $+0.053$ & $[+0.005, +0.102]$ & $+2.1$ to $+3.4\%$ \\
 & $\delta_H$ & $+0.075$ & $[+0.030, +0.122]$ & $+3.0$ to $+4.8\%$ \\
 & $\delta_Q$ & $+0.123$ & $[+0.017, +0.250]$ & $+5.0$ to $+7.9\%$ \\
\midrule
Comments & $\delta_I$ & $-0.065$ & $[-0.129, -0.008]$ & $-5.7\%$ \\
 & $\delta_H$ & $-0.007$ & $[-0.062, +0.041]$ & $-0.6\%$ \\
 & $\delta_Q$ & $+0.067$ & $[-0.062, +0.198]$ & $+5.9\%$ \\
\midrule
Comments & $\delta_I$ & $-0.0007$ & $[-0.0044, +0.0033]$ & $-0.5$ to $-0.8\%$ \\
(Binoculars) & $\delta_H$ & $+0.0014$ & $[-0.0014, +0.0043]$ & $+0.9$ to $+1.5\%$ \\
 & $\delta_Q$ & $+0.0023$ & $[-0.0063, +0.0107]$ & $+1.6$ to $+2.5\%$ \\
\bottomrule
\end{tabular}
\caption{Detector difference-in-differences for posts and comments, and the implied change in the share of machine-written documents relative to controls. Post shares give the range across the four calibrations (separations $+2.48$ to $+1.56$). Fast-DetectGPT comment shares use the GPT-4o calibration ($+1.15$), the only comment calibration with enough separation to use (Section~7.3); Binoculars comment rows are on that detector's own score scale, signed so that positive means more machine-like, with shares given as the range across its two usable calibrations (GPT-4o and Claude Sonnet 4). The post-side threshold of 5.1\% is the share of AI-written posts at which the volume interval would no longer rule out the 8.3\% decline.}
\label{tab:detector}
\end{table}

\subsection{Reddit comments show no differential rise in machine-likeness}

While the questions asked on Reddit seem to remain human-written, substitution may also appear in the answers, the comments under each post. We therefore test whether comments show any rise in machine-generated text after ChatGPT's release. We use the identical design and calibration with 223{,}775 comments from the 149 communities of the long-window sample, scored by both detectors.

Under Fast-DetectGPT, calibrated on GPT-4o (the only LLM it can reliably detect in comments), the estimate is negative. Machine-written comments fell by 5.7\% (CI $[-11.2\%, -0.7\%]$) in informational communities relative to controls. Binoculars, which can reliably detect both GPT-4o and Claude Sonnet 4 replies, does not reproduce this decline. Its estimate is instead a tight null, a decrease of 0.5\% (CI $[-3.0\%, +2.2\%]$) under the GPT-4o calibration and 0.8\% (CI $[-4.8\%, +3.7\%]$) under Claude Sonnet 4 (Table~\ref{tab:detector}). In other words, machine-generated comments did not accumulate faster in informational communities than in controls, and the answers people receive show no shift toward machine-generated content.

\section{Discussion}

While several of our community-type comparisons are dominated by pre-existing trends, we can still make three readings of our findings. First, the substitutable questions may already have left. The communities built on exactly the questions a model answers well, group Q, were losing a fifth of their volume every year before the launch, plausibly to search itself; ChatGPT arrived after those questions had already been leaving for years. Second, a public question does work a private chat does not. It leaves behind an archived, searchable answer, judgment from people who have been in the asker's position, and accountability that a model's fluent answer lacks. The question-asking literature has long observed that answers in these communities are valued for being from someone, not just for being correct \citep{anderson2012,cutrona1992}. Third, AI can substitute and help at the same time. A model helps a user draft a better question or rules out easy answers before they post, which raises the value of the questions that remain. Our design cannot separate these mechanisms, but each is consistent with the central finding that people did not stop asking.

Our claims in this study are limited to Reddit. Reddit's help-seeking communities use pseudonyms, are publicly archived, and show up in search, meaning a question asked there helps more people than its asker, and the platform's answers were part of what search engines and, later, language models learned from \citep{vincent2019}. Substitution may behave differently where those properties fail, such as private groups whose answers are not archived, identity-bound networks where asking costs reputation, or expert platforms like Stack Overflow, whose stricter posting rules mean a borderline question may be taken to a model instead. Under this reading, our null and the declines reported on Stack Overflow \citep{delrio2024,xue2023} reflect differences between platforms rather than contradictory estimates. Within Reddit, our sample is English-language communities above a volume floor; smaller and non-English communities are unmeasured.

The methods half of the paper applies beyond this study. Platform-scale generative-AI effects are being estimated across Q\&A sites, encyclopedias, and forums, usually as before-and-after contrasts within the affected population \citep{delrio2024,gaohahn2025,burtch2024,lyu2025}. Our placebo series shows that on Reddit the community types being compared move apart by up to 0.3 log points a year with no event at all, which is larger than most effects under study. We therefore recommend two inexpensive checks for future studies of platform-scale events. The first is a contemporaneous control group on the same platform with far less plausible exposure to the event, and the second is the same regression run at earlier dates where the answer is known to be zero. Both are available to any study using archival platform data. Finally, we recommend that future studies test whether machine-generated content inflates their counts, since any count of posts after 2023 may include it.

\section{Limitations}
\label{sec:limits}

The direct comparison between informational and human-anchored communities ($\Delta$) cannot be interpreted in our data. Its placebo estimates move over a range of 0.41 log points, and the drift is a property of group~H communities generally, not of the particular ones we sampled. We therefore rest no conclusion on it. Our main estimate instead compares each group with contemporaneous controls, a comparison that passes the placebo checks $\Delta$ fails.

Before seeing the data, the smallest decline our design could reliably detect was about 14\%; the realized confidence interval is tighter and rules out declines beyond 3.4\%, including the 8\% and larger estimates of prior work. It cannot resolve the 2.6\% reported in \citeauthor{xue2023}'s conference version, and our null is not evidence against an effect of that size. The interval still separates our result from every larger published estimate, and a decline of a few percent would describe a modest adjustment rather than the substitution the literature reports.

The primary window ends in May 2023, before the June 2023 API-pricing changes and the coordinated subreddit blackouts that followed, which keeps those events out of the estimate but limits it to the first six months after launch. The long window extends coverage to late 2025 at the cost of the overlapping events described in Section~\ref{sec:longwindow}, and we do not attempt to divide its $+24\%$ among search-driven visibility, machine-generated content, and real demand. The two windows together still cover both ends of the period, as no decline appears in the first six months and none emerges once adoption reaches scale.

The detector analysis has four stated limits. First, the calibration rests on four LLM families whose post-side separations span $+1.56$ to $+2.48$, which brackets but does not remove the dependence on the LLM; real machine-generated text on Reddit is a broader mixture including human-edited output, and text engineered to dodge statistical detectors would fall outside all four calibrations. Second, the share of documents long enough to score rose on the post side between the periods, and rose about twice as steeply in controls as in group~I (Appendix~D); holding each community's pre-period length mix fixed leaves the estimates unchanged, but posts too short to score in either period stay outside the analysis. Third, body-conditioned replies from DeepSeek-V3 and GLM-5.2 fall below the separation the conversion requires for both detectors; the comment analysis therefore rests on the GPT-4o and Claude Sonnet 4 calibrations (through Binoculars) and bounds only machine-generated text that some detector can see; answers written with the least detectable models are invisible to both detectors. Fourth, detector scores are documented to vary with writer population \citep{liang2023}; the design cancels stable bias, but a change in who writes within a group over the window would register as a score shift. For all these reasons the detector numbers are limits and consistency checks, not estimates of how much machine-generated content exists, and that is the only use we put them to; the volume results stand on counts alone and do not depend on them.

\bibliographystyle{aaai2026}
\bibliography{refs}

\begin{thebibliography}{29}
\providecommand{\natexlab}[1]{#1}

\bibitem[{Anderson et~al.(2012)Anderson, Huttenlocher, Kleinberg, and
  Leskovec}]{anderson2012}
Anderson, A.; Huttenlocher, D.; Kleinberg, J.; and Leskovec, J. 2012.
\newblock Discovering value from community activity on focused question
  answering sites: a case study of {S}tack {O}verflow.
\newblock In \emph{Proceedings of the 18th ACM SIGKDD International Conference
  on Knowledge Discovery and Data Mining}, 850--858.

\bibitem[{Bao et~al.(2024)Bao, Zhao, Teng, Yang, and Zhang}]{bao2024}
Bao, G.; Zhao, Y.; Teng, Z.; Yang, L.; and Zhang, Y. 2024.
\newblock Fast-{DetectGPT}: Efficient Zero-Shot Detection of Machine-Generated
  Text via Conditional Probability Curvature.
\newblock In \emph{Proceedings of the Twelfth International Conference on
  Learning Representations (ICLR)}.

\bibitem[{Baumgartner et~al.(2020)Baumgartner, Zannettou, Keegan, Squire, and
  Blackburn}]{baumgartner2020}
Baumgartner, J.; Zannettou, S.; Keegan, B.; Squire, M.; and Blackburn, J. 2020.
\newblock The {P}ushshift {R}eddit Dataset.
\newblock \emph{Proceedings of the International AAAI Conference on Web and
  Social Media}, 14: 830--839.

\bibitem[{Bertrand, Duflo, and Mullainathan(2004)}]{bertrand2004}
Bertrand, M.; Duflo, E.; and Mullainathan, S. 2004.
\newblock How Much Should We Trust Differences-In-Differences Estimates?
\newblock \emph{The Quarterly Journal of Economics}, 119(1): 249--275.

\bibitem[{Bick, Blandin, and Deming(2024)}]{bick2024}
Bick, A.; Blandin, A.; and Deming, D.~J. 2024.
\newblock The Rapid Adoption of Generative {AI}.
\newblock Working Paper 32966, National Bureau of Economic Research.

\bibitem[{Black et~al.(2021)Black, Gao, Wang, Leahy, and Biderman}]{gptneo2021}
Black, S.; Gao, L.; Wang, P.; Leahy, C.; and Biderman, S. 2021.
\newblock {GPT}-{N}eo: Large Scale Autoregressive Language Modeling with
  Mesh-Tensorflow.
\newblock Zenodo.

\bibitem[{Burtch, Lee, and Chen(2024)}]{burtch2024}
Burtch, G.; Lee, D.; and Chen, Z. 2024.
\newblock The consequences of generative {AI} for online knowledge communities.
\newblock \emph{Scientific Reports}, 14: 10413.

\bibitem[{Cameron, Gelbach, and Miller(2008)}]{cameron2008}
Cameron, A.~C.; Gelbach, J.~B.; and Miller, D.~L. 2008.
\newblock Bootstrap-Based Improvements for Inference with Clustered Errors.
\newblock \emph{Review of Economics and Statistics}, 90(3): 414--427.

\bibitem[{Cutrona and Suhr(1992)}]{cutrona1992}
Cutrona, C.~E.; and Suhr, J.~A. 1992.
\newblock Controllability of Stressful Events and Satisfaction With Spouse
  Support Behaviors.
\newblock \emph{Communication Research}, 19(2): 154--174.

\bibitem[{De~Choudhury and De(2014)}]{dechoudhury2014}
De~Choudhury, M.; and De, S. 2014.
\newblock Mental Health Discourse on reddit: Self-Disclosure, Social Support,
  and Anonymity.
\newblock \emph{Proceedings of the International AAAI Conference on Web and
  Social Media}, 8(1): 71--80.

\bibitem[{del Rio-Chanona, Laurentsyeva, and Wachs(2024)}]{delrio2024}
del Rio-Chanona, R.~M.; Laurentsyeva, N.; and Wachs, J. 2024.
\newblock Large language models reduce public knowledge sharing on online
  {Q}\&{A} platforms.
\newblock \emph{PNAS Nexus}, 3(9): pgae400.

\bibitem[{Gao and Hahn(2025)}]{gaohahn2025}
Gao, Y.; and Hahn, J. 2025.
\newblock The Impact of {ChatGPT} on People's Engagement with Online Advice
  Communities.
\newblock In \emph{Proceedings of the 58th Annual Hawaii International
  Conference on System Sciences (HICSS-58)}.

\bibitem[{Hans et~al.(2024)Hans, Schwarzschild, Cherepanova, Kazemi, Saha,
  Goldblum, Geiping, and Goldstein}]{hans2024}
Hans, A.; Schwarzschild, A.; Cherepanova, V.; Kazemi, H.; Saha, A.; Goldblum,
  M.; Geiping, J.; and Goldstein, T. 2024.
\newblock Spotting {LLMs} With Binoculars: Zero-Shot Detection of
  Machine-Generated Text.
\newblock In \emph{Proceedings of the 41st International Conference on Machine
  Learning (ICML), PMLR 235}, 17519--17537.

\bibitem[{Koonchanok(2026)}]{koonchanok2026}
Koonchanok, R. 2026.
\newblock Do Students Still Ask Each Other? {E}vidence from {R}eddit on Test
  Preparation in the Age of Generative {AI}.
\newblock \emph{Proceedings of the International AAAI Conference on Web and
  Social Media}, 20(1): 2996--3001.

\bibitem[{Liang et~al.(2023)Liang, Yuksekgonul, Mao, Wu, and Zou}]{liang2023}
Liang, W.; Yuksekgonul, M.; Mao, Y.; Wu, E.; and Zou, J. 2023.
\newblock {GPT} detectors are biased against non-native {E}nglish writers.
\newblock \emph{Patterns}, 4(7): 100779.

\bibitem[{Lyu et~al.(2025)Lyu, Siderius, Li, Acemoglu, Huttenlocher, and
  Ozdaglar}]{lyu2025}
Lyu, L.; Siderius, J.; Li, H.; Acemoglu, D.; Huttenlocher, D.; and Ozdaglar, A.
  2025.
\newblock Wikipedia Contributions in the Wake of {ChatGPT}.
\newblock In \emph{Companion Proceedings of the ACM on Web Conference 2025 (WWW
  '25 Companion)}, 1176--1179.

\bibitem[{MacKinnon and Webb(2018)}]{mackinnonwebb2018}
MacKinnon, J.~G.; and Webb, M.~D. 2018.
\newblock The wild bootstrap for few (treated) clusters.
\newblock \emph{The Econometrics Journal}, 21(2): 114--135.

\bibitem[{Mitchell et~al.(2023)Mitchell, Lee, Khazatsky, Manning, and
  Finn}]{mitchell2023}
Mitchell, E.; Lee, Y.; Khazatsky, A.; Manning, C.~D.; and Finn, C. 2023.
\newblock {DetectGPT}: Zero-Shot Machine-Generated Text Detection using
  Probability Curvature.
\newblock In \emph{Proceedings of the 40th International Conference on Machine
  Learning (ICML)}.

\bibitem[{Proferes et~al.(2021)Proferes, Jones, Gilbert, Fiesler, and
  Zimmer}]{proferes2021}
Proferes, N.; Jones, N.; Gilbert, S.; Fiesler, C.; and Zimmer, M. 2021.
\newblock Studying {R}eddit: A Systematic Overview of Disciplines, Approaches,
  Methods, and Ethics.
\newblock \emph{Social Media + Society}, 7(2).

\bibitem[{Quinn and Gutt(2025)}]{quinngutt2025}
Quinn, M.; and Gutt, D. 2025.
\newblock Heterogeneous Effects of Generative Artificial Intelligence
  ({G}en{AI}) on Knowledge Seeking in Online Communities.
\newblock \emph{Journal of Management Information Systems}, 42(2): 370--399.

\bibitem[{Rambachan and Roth(2023)}]{rambachanroth2023}
Rambachan, A.; and Roth, J. 2023.
\newblock A More Credible Approach to Parallel Trends.
\newblock \emph{Review of Economic Studies}, 90(5): 2555--2591.

\bibitem[{Roth(2022)}]{roth2022}
Roth, J. 2022.
\newblock Pretest with Caution: Event-Study Estimates after Testing for
  Parallel Trends.
\newblock \emph{American Economic Review: Insights}, 4(3): 305--322.

\bibitem[{Shan and Qiu(2026)}]{shanqiu2025}
Shan, G.; and Qiu, L. 2026.
\newblock Examining the Impact of Generative {AI} on Users' Voluntary Knowledge
  Contribution: Evidence from a Natural Experiment on {S}tack {O}verflow.
\newblock \emph{Information Systems Research}, 37(2): 1021--1041.

\bibitem[{Vincent et~al.(2019)Vincent, Johnson, Sheehan, and
  Hecht}]{vincent2019}
Vincent, N.; Johnson, I.; Sheehan, P.; and Hecht, B. 2019.
\newblock Measuring the Importance of User-Generated Content to Search Engines.
\newblock \emph{Proceedings of the International AAAI Conference on Web and
  Social Media}, 13: 505--516.

\bibitem[{Wester et~al.(2024)Wester, Schrills, Pohl, and van
  Berkel}]{wester2024}
Wester, J.; Schrills, T.; Pohl, H.; and van Berkel, N. 2024.
\newblock ``As an {AI} language model, {I} cannot'': Investigating {LLM}
  Denials of User Requests.
\newblock In \emph{Proceedings of the CHI Conference on Human Factors in
  Computing Systems}, 1--14.

\bibitem[{Xue et~al.(2023)Xue, Wang, Zheng, Li, and Tan}]{xue2023}
Xue, J.; Wang, L.; Zheng, J.; Li, Y.; and Tan, Y. 2023.
\newblock {ChatGPT} Is A User-Generated Knowledge-Sharing Killer.
\newblock In \emph{Proceedings of the International Conference on Information
  Systems (ICIS)}.

\bibitem[{Xue et~al.(2026)Xue, Wang, Zheng, Li, and Tan}]{xue2026}
Xue, J.; Wang, L.; Zheng, J.; Li, Y.; and Tan, Y. 2026.
\newblock Can {ChatGPT} Kill User-Generated {Q}\&{A} Platforms?
\newblock \emph{Information Systems Research}.
\newblock Articles in Advance.

\bibitem[{Yin, Jia, and Wakslak(2024)}]{yin2024}
Yin, Y.; Jia, N.; and Wakslak, C.~J. 2024.
\newblock {AI} can help people feel heard, but an {AI} label diminishes this
  impact.
\newblock \emph{Proceedings of the National Academy of Sciences}, 121(14):
  e2319112121.

\bibitem[{Zhang, Cui, and Zhang(2026)}]{zhang2026}
Zhang, P.; Cui, R.; and Zhang, D.~J. 2026.
\newblock The Impact of {AI} Search on the Online Content Ecosystem: Evidence
  from {Google} and {Reddit}.
\newblock \emph{arXiv preprint arXiv:2605.16428}.

\end{thebibliography}

\clearpage
\appendix

\section*{Ethics Statement}

The counting analysis uses aggregate monthly counts of public posts; no text, comment, or author identifier enters it. The detector analysis reads sampled public post and comment text under a deliberately narrow scope. The author field is never requested, deleted and removed content is dropped, text is scored by machine and no member of the research team reads or quotes any document, and scoring outputs retain only (group, community, month, length, score). Text leaves the compute environment where it is scored in one scoped case. Generating the calibration corpus transmitted the public titles of 300 pre-period posts from groups I and H, and for the comment calibration their bodies truncated to 1{,}500 characters, to four commercial model APIs through a gateway; no author information accompanied any prompt. Group H contains bereavement, abuse-recovery and crisis communities, which is why the pipeline is built so that participation in them is never read by a member of the research team or excerpted in any artifact. All reported quantities are aggregates over hundreds of communities; no individual-level record is constructed and no claim is made about any individual or any single community's members. Detector scores are never used to label any document, author, or community as machine-generated; they enter only as group-level means in a differencing design that cancels stable detector bias. We report the documented population biases of such detectors \citep{liang2023} as a limitation. The study is secondary analysis of public data and involves no interaction with human subjects; our handling follows the norms documented for Reddit research \citep{proferes2021}, and data use follows the archive's terms and Reddit's user agreement.

\section{Placebo Construction Robustness}
\label{app:placebo}

This appendix reports the 66-window placebo sweep in full and tests two objections to it.

\paragraph{The sweep in full.} Every calendar anchor that allows a seasonally matched six-versus-six-month window ending before ChatGPT's release contributes one run, 66 in all.

\begin{table}[h]
\centering\small
\begin{tabular}{@{}lrrrr@{}}
\toprule
 & mean & SD & range & \% negative \\
\midrule
$\delta_I$ & $-0.004$ & 0.062 & $[-0.133, +0.093]$ & 65\% \\
$\delta_H$ & $+0.104$ & 0.122 & $[-0.187, +0.242]$ & 18\% \\
$\delta_Q$ & $-0.197$ & 0.045 & $[-0.285, -0.108]$ & 100\% \\
$\Delta$ & $-0.108$ & 0.123 & $[-0.245, +0.162]$ & 80\% \\
\bottomrule
\end{tabular}
\caption{All 66 placebo estimates, in log points of monthly posts relative to controls; the true effect at each date is zero. Each mean shows that quantity's pre-existing trend; $\delta_I$ is the only one of the four whose mean is close to zero.}
\end{table}

\paragraph{The windows overlap.} Windows anchored a month apart share ten of their twelve months, and the 66 runs are therefore far from independent draws. The overlap does not bias the mean, which is what our choice of estimate rests on. It does mean the SD above describes the spread of a correlated family rather than a sampling distribution, and we use it descriptively. Restricting to the five December anchors, which share only six months with their neighbors, leaves the same picture ($\delta_I$ mean $+0.011$, $\Delta$ mean $-0.100$), as does splitting those into the non-overlapping subsets \{2017, 2019, 2021\} ($+0.020$, $-0.124$) and \{2018, 2020\} ($-0.003$, $-0.063$).

\paragraph{Sample composition might drift across windows.} A window enters only if a community reports every one of its twelve months, and composition could therefore in principle move with the anchor; in practice it barely does. The full sample of 26/29/6/90 is present in 53 of the 66 windows, and no window loses more than three communities, all of them archive gaps rather than dormancy. The one recurring exception, r/CasualUK, is genuinely missing 2016-12 and 2017-01 and is dropped from the windows that need those months rather than entering as a zero.

\section{Estimate and Window Diagnostics}
\label{app:estimand}

All quantities in this appendix are computed on pre-ChatGPT data only.

\paragraph{Why $\delta_I$ rather than $\Delta$.} The between-group contrast $\Delta = \delta_I - \delta_H$ is significantly non-zero at three of the five December-anchor placebo dates, the near-independent subset of the 66-window sweep. All three are negative, and the nearest to the launch is $-0.142$, 94\% of the target effect size. $\delta_I$ passes every placebo check, with mean $+0.011$; its across-placebo standard deviation (0.052) matches its within-window clustered standard error (0.048; ratio 1.09); community-clustered inference is therefore consistent with the placebo dispersion.

\paragraph{Alternative estimators do not repair $\Delta$.} We benchmarked four candidate estimators on placebo bias and variance: the two-way FE estimator of the main text, the same estimator with group-specific linear pre-trends, a matched-pairs difference on community-level year-over-year changes, and a rank-based version robust to outlier months. Every variant leaves $\Delta$'s placebo bias at or above two-thirds of the target effect size, because the drift is real heterogeneity across communities rather than a functional-form artifact. Group-specific linear detrending mechanically forces placebo means toward zero in-sample and was rejected for extrapolation risk under \citet{roth2022} and \citet{rambachanroth2023}.

\paragraph{Seasonality.} Over 2018--2021, group~I's December-to-May months average $-0.008$ log points relative to the group's annual mean, group~H's $-0.047$, and group~C's $-0.006$. An unmatched window therefore biases the I-versus-H contrast by roughly 4\% while leaving group-versus-control contrasts nearly seasonally neutral; the seasonally matched window removes the seasonal term entirely.

\paragraph{Window-length dispersion.} The cross-community standard deviation of the pre-to-post shift in $\log(1+y)$, holding the six-month shape fixed and moving the post window out from the pre window, rises monotonically from 0.310 (3-month separation) to 0.349 (24-month separation). Under month-level noise this dispersion would shrink with averaging distance. Its growth identifies persistent community-level drift as the dominant error component, which motivates both the near post-window and community-level clustering.

\section{The Community Sample}
\label{app:frame}

The sample was assigned before any outcome was computed, then filtered by four mechanical rules on pre-ChatGPT data (founding date on or before 2017-12; median monthly volume $\ge 100$; $\ge 50$ posts in 90\% of months; no missing months). Six candidates failed (r/AskALawyer and r/DadForAMinute on founding date, r/whittling on the volume floor, and r/AskDentists, r/FirstTimeHomeBuyer and r/crossword on coverage).

\paragraph{Group I, informational advice (26).} AskCarSales, AskDocs, AskProgramming, AskVet, CreditCards, DIY, HomeImprovement, Insurance, LegalAdviceUK, MechanicAdvice, PersonalFinanceCanada, RealEstate, StudentLoans, UKPersonalFinance, buildapc, careerguidance, excel, immigration, learnprogramming, learnpython, legaladvice, personalfinance, resumes, smallbusiness, tax, techsupport.

\paragraph{Group H, human-anchored support (29).} AlAnon, Anxiety, BipolarReddit, BreakUps, CPTSD, ExNoContact, ForeverAlone, GriefSupport, KindVoice, MomForAMinute, SuicideWatch, TrueOffMyChest, adultsurvivors, beyondthebump, confession, dating\_advice, depression, divorce, leaves, lonely, offmychest, ptsd, relationship\_advice, relationships, self, socialanxiety, stepparents, stopdrinking, widowers.

\paragraph{Group Q, low-stakes curiosity (6).} AskHistorians, answers, askscience, explainlikeimfive, tipofmytongue, whatisthisthing.

\paragraph{Group C, controls (90).} 3Dprinting, Aquariums, Astronomy, Bass, BeardedDragons, Blacksmith, Bonsai, Breadit, Bushcraft, Canning, CasualUK, Charcuterie, Cichlid, Coffee, Cooking, DnD, EDC, Embroidery, Fantasy, Fishing, Genealogy, Guitar, Homebrewing, Kayaking, Leathercraft, MartialArts, Mountaineering, Origami, Pizza, PlantedTank, Rabbits, Sneakers, Ultralight, VintageComputing, Warhammer40k, Watches, WeAreTheMusicMakers, analog, backpacking, baking, battlestations, bicycling, birding, boardgames, books, bouldering, budgies, calligraphy, campingandhiking, castiron, chess, climbing, cocktails, comicbooks, crochet, cycling, drums, fermentation, flyfishing, fountainpens, geocaching, grilling, hiking, houseplants, jigsawpuzzles, knitting, knives, lego, magicTCG, mechanicalkeyboards, metalworking, museum, mycology, orchids, photography, piano, printSF, quilting, reptiles, running, scifi, sewing, shrimptank, sousvide, succulents, suggestmeabook, synthesizers, tea, vinyl, woodworking.

Two group-I communities (DIY, HomeImprovement) lack complete coverage over the extended 2018--2025 panel and drop from the constant long-window sample, which is therefore 24~I / 29~H / 6~Q / 90~C.

\section{Shares of Documents Long Enough to Score}
\label{app:textshare}

The detector scores only documents of at least 200 characters. The tables report the share of sampled documents passing that floor by group and period, which is the composition check the exclusion requires. They cover the full sampled corpus of all 157 candidate communities, the sample in which selection into the floor is measured; the difference-in-differences itself uses the constant sample of Section~\ref{sec:longwindow} (274{,}411 post and 223{,}775 comment scores).

\begin{table}[h]
\centering\small
\begin{tabular}{@{}lrrrrr@{}}
\toprule
Group & pre share & post share & pre $n$ & post $n$ & subs \\
\midrule
I & 0.536 & 0.746 & 35{,}698 & 49{,}549 & 29/29 \\
H & 0.485 & 0.705 & 33{,}434 & 48{,}634 & 30/30 \\
Q & 0.216 & 0.274 & 2{,}982 & 3{,}787 & 6/6 \\
C & 0.207 & 0.421 & 39{,}606 & 80{,}229 & 83/92 \\
\bottomrule
\end{tabular}
\caption{Posts. The share passing the length floor rises in every group pre-to-post, by about half in the treated groups and doubling in controls, a composition shift we discuss in the Limitations section. Counts are documents long enough to score.}
\end{table}

\begin{table}[h]
\centering\small
\begin{tabular}{@{}lrrrrr@{}}
\toprule
Group & pre share & post share & pre $n$ & post $n$ & subs \\
\midrule
I & 0.472 & 0.436 & 29{,}095 & 27{,}801 & 29/29 \\
H & 0.454 & 0.448 & 28{,}316 & 29{,}649 & 30/30 \\
Q & 0.472 & 0.474 & 5{,}147 & 5{,}775 & 6/6 \\
C & 0.267 & 0.272 & 54{,}323 & 56{,}827 & 92/92 \\
\bottomrule
\end{tabular}
\caption{Comments. Shares are essentially flat pre-to-post in every group; the post-side composition shift does not extend to comments.}
\end{table}

Ten long-window control communities have no posts long enough to score in one of the two periods and drop from the post-side detector sample (C $= 80$ there); all 90 qualify on the comment side.

\section{Detector Calibration Details}
\label{app:calib}
\label{app:detector}

\paragraph{Scoring.} Fast-DetectGPT conditional probability curvature \citep{bao2024} with GPT-Neo-1.3B \citep{gptneo2021} as the scoring model, fp16, 512-token truncation, documents batched at 32. All scoring ran on a single NVIDIA A100 on an institutional cluster, roughly 2.5 GPU-hours in total; generation used a commercial API gateway at a cost under ten US dollars. Scores are computed once per document; the analysis consumes per-document $(g, s, m, n_{\mathrm{body}}, d)$ tuples only.

\paragraph{Positive class.} 300 posts sampled uniformly (seed 0) from pre-period group-I/H posts that pass the length floor. For each, the four LLMs (GPT-4o, Claude Sonnet 4, DeepSeek-V3, and GLM-5.2; temperature 0.9, max 1024 tokens) produce a matching document. The post prompt is ``Write the body of a Reddit post for r/\emph{sub} titled: \emph{title}. First person, 250--450 words, plain casual prose as a real redditor would write it. Output only the post body, nothing else.'' The comment prompt conditions on the real post: ``Write a Reddit comment replying to this post on r/\emph{sub}: Title: \emph{title} \emph{body} Give practical advice or an answer as a real redditor would, 120--300 words, plain casual prose. Output only the comment, nothing else.'', with the body truncated at 1{,}500 characters. Success rates were 300/300 for both document types for the first three LLMs, and 287/300 (posts) and 297/300 (comments) for GLM-5.2 (transient API failures). Of the successes, all passed the 200-character floor except two GLM-5.2 post generations (285 scored). All were scored by the identical pipeline.

\paragraph{Calibration outcomes.} All AUROCs are rank comparisons with ties half-weighted, against a matched-community pre-period human mean of $-0.112$ (SD 1.020, $n = 68{,}260$) for posts and $-0.236$ (SD 1.054, $n = 56{,}329$) for comments. Posts, generated means: GPT-4o $+2.370$ (SD 0.901), separation $+2.48$, AUROC 0.966; Claude Sonnet 4 $+1.771$ (SD 0.921), $+1.88$, 0.916; DeepSeek-V3 $+1.640$ (SD 1.041), $+1.75$, 0.889; GLM-5.2 $+1.443$ (SD 0.915), $+1.56$, 0.875. Comments: GPT-4o $+0.913$ (SD 0.916), separation $+1.15$, AUROC 0.794; Claude Sonnet 4 $+0.337$ (SD 0.880), $+0.57$, 0.664; DeepSeek-V3 $+0.082$ (SD 1.028), $+0.32$, 0.585; GLM-5.2 $-0.036$ (SD 0.947), $+0.20$, 0.552.

\paragraph{Second detector.} Binoculars \citep{hans2024} scores each document as the ratio of an observer model's perplexity to the cross-perplexity between an observer and a performer (tiiuae/falcon-7b and falcon-7b-instruct), fp16, 512-token truncation, with \emph{lower} scores indicating machine-generated text. We report its estimates signed so that positive means more machine-like. On a 5{,}000-document human baseline per document type, its post AUROCs against the same four LLMs are 0.967 (GPT-4o), 0.963 (Claude Sonnet 4), 0.935 (DeepSeek-V3), and 0.896 (GLM-5.2), confirming the Fast-DetectGPT post calibration. On comments, against the full pre-period human baseline (mean 0.9833, SD 0.076, $n = 56{,}329$), generated means are 0.8351 (GPT-4o, AUROC 0.939, raw separation 0.148), 0.8921 (Claude Sonnet 4, 0.850, 0.091), 0.9160 (DeepSeek-V3, 0.768, 0.067), and 0.9360 (GLM-5.2, 0.698, 0.047). The first two separations support conversion and the last two do not. Rescoring all sampled comments and re-running the identical difference-in-differences on the constant sample yields the Binoculars rows of Table~\ref{tab:detector}. Binoculars scoring used roughly 5.5 additional A100 GPU-hours.

\paragraph{Conversion.} If a fraction $f$ of a group's documents is machine-generated with mean score elevated by the separation $S$, the group's mean score shifts by $fS$; a differential DiD shift of $\delta$ therefore corresponds to a differential share change of $\delta / S$. The conversion assumes post-period machine-generated text scores like the calibration text; the sensitivity analysis in the main text degrades $S$ threefold before the conclusion for posts changes.

\paragraph{Within-group shifts.} All four groups' mean post scores fall from pre to post (I $-0.102$, H $-0.079$, Q $-0.031$, C $-0.155$), as do comment scores (I $-0.138$, H $-0.080$, Q $-0.005$, C $-0.073$). A level drift common to all groups (vocabulary change, platform composition, archive coverage) is exactly what the differencing removes; no level is interpreted.

\paragraph{Length composition.} The share of posts long enough to score rises over the window, and by more in controls than in group~I (Appendix~\ref{app:textshare}); a period mean could therefore move because its length mix moved. Averaging within-bin means under each community's own pre-period length weights gives $\delta_I = +0.055$ $[+0.013, +0.099]$, $\delta_H = +0.081$ and $\delta_Q = +0.154$, against $+0.053$, $+0.075$ and $+0.123$ unadjusted; pooled pre-period deciles instead of per-community weights give $\delta_I = +0.052$. Restricting to documents of at least 500 characters selects a different population rather than adjusting composition. It gives a larger $\delta_I = +0.087$ $[+0.030, +0.148]$, which converts to $3.5\%$ to $5.6\%$ and reaches the 5.1\% threshold under the weakest calibration.

\paragraph{Sensitivity of the conversion to the separation.} The implied differential share is $\delta / S$; any overstatement of the separation $S$ therefore understates inflation proportionally.

\begin{table}[h]
\centering\small
\begin{tabular}{@{}lrr@{}}
\toprule
Assumed $S$ & implied $\delta_I$ share (CI) & vs.\ 5.1\% bar \\
\midrule
$2.48$ (GPT-4o) & $+2.1\%$ $[0.2, 4.1]$ & under \\
$1.88$ (Claude Sonnet 4) & $+2.8\%$ $[0.2, 5.4]$ & point under \\
$1.75$ (DeepSeek-V3) & $+3.0\%$ $[0.3, 5.8]$ & point under \\
$1.56$ (GLM-5.2) & $+3.4\%$ $[0.3, 6.5]$ & point under \\
$1.04$ & $+5.1\%$ & threshold \\
\bottomrule
\end{tabular}
\caption{Conversions under each measured calibration, as the implied differential share of posts. All four LLMs leave the point estimate under the threshold at which the 8.3\% decline is no longer ruled out; only under the strongest (GPT-4o) does the interval also rule it out. An LLM whose separation falls below 1.04 on this text would be invisible to the test.}
\end{table}

\section{The Long Window in Full}
\label{app:longwindow}

The constant sample over 2018--2025 is 24~I / 29~H / 6~Q / 90~C. Window B is the specification of Section~\ref{sec:longwindow}; window A is the same post-period with the excluded months absorbed into the pre-period (its pre-period then contains twelve months of ChatGPT and the API blackout, attenuating estimates toward zero).

\begin{table}[h]
\centering\small
\begin{tabular}{@{}lrrrr@{}}
\toprule
 & $\delta_I$ & $\delta_H$ & $\delta_Q$ & $\Delta$ \\
\midrule
A treatment & $+0.158$ & $+0.069$ & $-0.182$ & $+0.089$ \\
A placebo ($-3$y) & $+0.098$ & $+0.190$ & $-0.458$ & $-0.091$ \\
B treatment & $+0.218$ & $+0.098$ & $-0.251$ & $+0.120$ \\
B placebo ($-3$y) & $+0.092$ & $+0.188$ & $-0.561$ & $-0.096$ \\
B drift-adjusted & $+0.126$ & $-0.090$ & $+0.311$ & $+0.216$ \\
\bottomrule
\end{tabular}
\caption{Long-window estimates, in log points of monthly posts relative to controls. The $\delta_I$ estimate has $p = 0.040$. Only one placebo window exists at this length, and the drift adjustment therefore carries no interval; $\delta_Q$'s placebo estimate of $-0.561$ shows drift dominating at this horizon.}
\end{table}

\section{Reproducibility}
\label{app:repro}

Every number in the paper regenerates from scripts that read the archived count and score files. These cover the sample pull and counting analysis, the placebo and pre-trend battery, the estimator comparison, the long-window table, both detector difference-in-differences, and both calibrations. Each script carries a self-check mode that asserts the published cells against a fresh recomputation, and each figure regenerates from a single script reading the same intermediate files. Counts were pulled from the public archive endpoint with the query log retained; six randomly chosen cached counts were re-verified against the live endpoint during the final analysis pass. Code and intermediate aggregates (counts and per-document score tuples; no text) are available at \url{https://github.com/hazemibrahim97/informational-reddit-chatgpt}.

\end{document}